\documentclass[journal=cmatex]{achemso}
\usepackage[T1]{fontenc} 
\usepackage[version=3]{mhchem}
\usepackage{amsmath}
\usepackage{amssymb}
\usepackage{graphicx}
\usepackage{subcaption}
\usepackage{color}
\usepackage{dcolumn}
\usepackage{bm}
\usepackage{miller}
\usepackage[utf8]{inputenc}
\usepackage{nicefrac}
\usepackage{multirow}
\usepackage{makecell}
\usepackage{nameref}
\usepackage{hyperref}
\usepackage{pdfpages}
\hypersetup{
    colorlinks=true,
    linkcolor=blue,
    filecolor=magenta, 
    urlcolor=cyan,
    citecolor=blue
}

\definecolor{orange}{RGB}{255, 108, 12}

\graphicspath{ {Figures/} }
\author{Victor Venturi}
\affiliation{Department of Mechanical Engineering, Carnegie Mellon University, Pittsburgh, Pennsylvania 15213, USA}
\author{Venkatasubramanian Viswanathan}
\affiliation{Department of Mechanical Engineering, Carnegie Mellon University, Pittsburgh, Pennsylvania 15213, USA}
\alsoaffiliation[phys]{Department of Physics, Carnegie Mellon University, Pittsburgh, Pennsylvania 15213, USA}
\email{venkvis@cmu.edu}

\title[An \textsf{achemso} demo]
  {Thermodynamics of Lithium Stripping and Limits for Fast Discharge in Lithium Metal Batteries}

\begin{document}

\begin{tocentry}
\includegraphics[width=9cm]{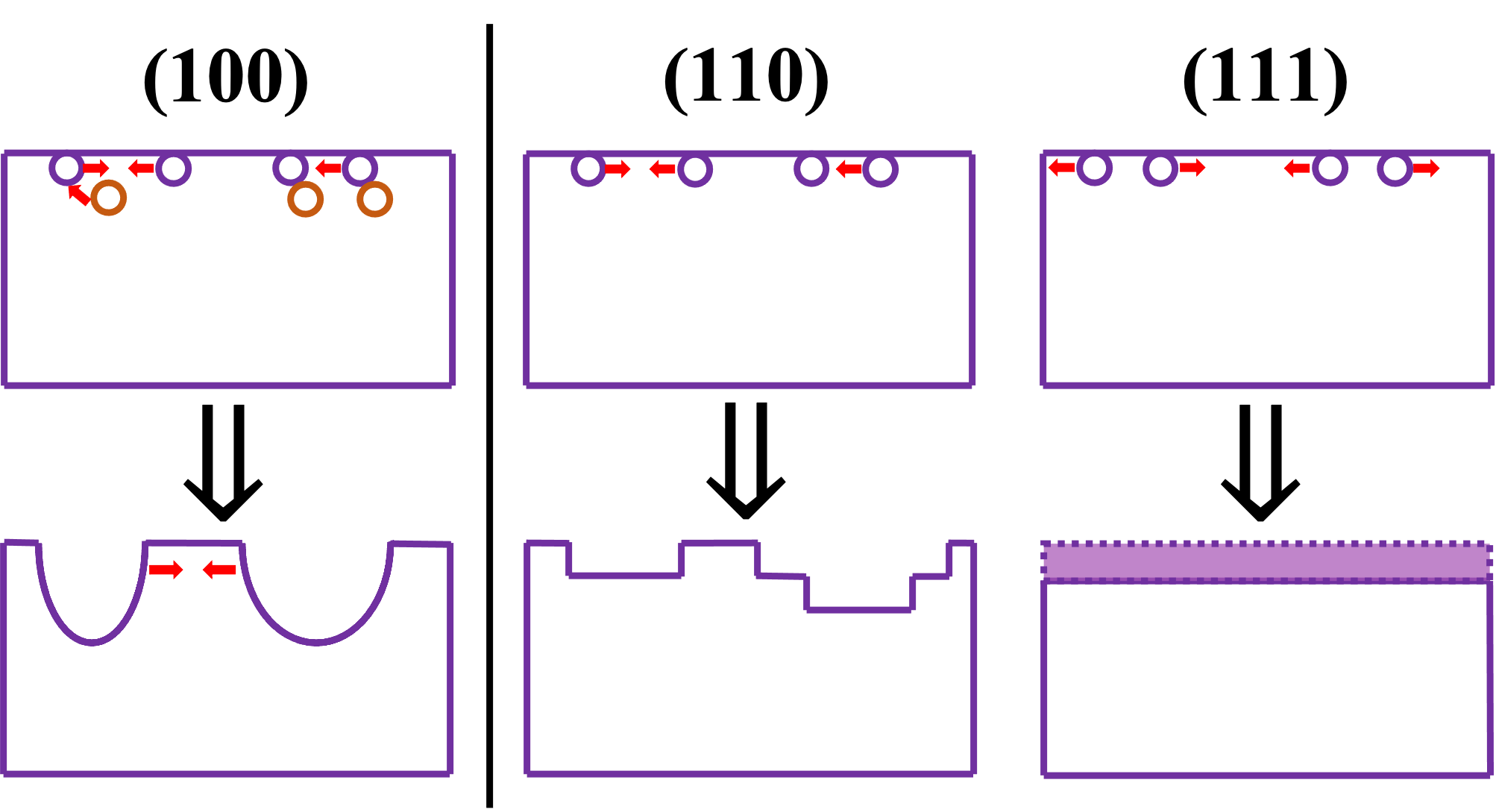}
\\
\end{tocentry}

\begin{abstract}
Lithium metal batteries are seen as a critical piece towards electrifying aviation. During charging, plating of lithium metal, a critical failure mechanism, has been studied and mitigation strategies have been proposed.  For electric aircraft, high discharge power requirements necessitate stripping of lithium metal in an uniform way and recent studies have identified the evolution of surface voids and pits as a potential failure mechanism. In this work, using density functional theory calculations and thermodynamic analysis, we investigate the discharge process on lithium metal surfaces.  In particular, we calculate the tendency for vacancy congregation on lithium metal surfaces, which constitutes the first step in the formation of voids and pits. We find that among the low Miller index surfaces, the (111) surface is the least likely to exhibit pitting issues. Our analysis suggests the faceting control during electrodeposition could be a key pathway towards simultaneously enabling both fast charge and fast discharge.
\end{abstract}




Lithium-ion batteries are now widely used for portable electronic devices and electric vehicles due to their large voltage window, high energy density, and versatility, among other advantageous characteristics~\cite{sun2016promises, aurbach2000factors, lin2017reviving}.  Electrification of long-haul trucks,\cite{sripad2017performance} vertical take-off and landing (eVTOL) aircraft,\cite{fredericks2018performance}  and regional and narrow-body aircraft,\cite{Epstein2019,bills2020universal} requires  higher specific energy than possible with current Li-ion batteries. Lithium metal anodes are seen as a critical enabler towards meeting these targets,\cite{xu2014lithium, lin2017reviving, fu2020universal} but technical hurdles remain hindering their commercialization.

Much of said issues stem from the myriad of processes that can take place at interfaces between different materials, but especially at the anode surface~\cite{bhowmik2019perspective, sharafi2017surface, exner2017constrained, exner2018short, gialampouki2019electrochemical}. During charging of the lithium metal battery, formation and growth of dendrites in the plating process has been widely-studied and identified as a critical failure mechanism and mitigation strategies have been proposed~\cite{yan2016selective, fu2020universal, li2015synergetic, lu2014stable, qian2015high, zhang2017lithium, zhao2017surface,  hatzell2020challenges, jackle2018self, jackle2014microscopic}. Electric aircraft applications, in particular, eVTOL, require high discharge power, which can cause new failure mechanisms during the lithium stripping process.~\cite{wang2019characterizing, wood2016dendrites, cohen2000micromorphological, shi2018lithium, gireaud2006lithium}. Since the morphology of the surface of the lithium anode post-discharge can heavily impact the deposition process in subsequent cycles~\cite{monroe2005impact, ahmad2017stability, kasemchainan2019critical}, it is critical to develop a fundamental understanding of void and pit formation associated with  lithium stripping. 

One of the earliest stages in the formation of these voids is the congregation of vacancies on the lithium surface. In order to understand this mechanism, we use a regular solution model to simulate the anode surface as a mixture of vacancies and lithium atoms, which are treated as two separate particles. The Gibbs free energy of mixing, normalized by the total number of particles, is given by
\begin{equation}
    \Delta g_{mix} = \frac{z}{2}x_V x_L(2\epsilon_{VL} - \epsilon_{VV} - \epsilon_{LL}) + k_BT\left[x_V \log\left(x_L\right) + x_V \log\left(x_L\right)\right],
\end{equation}
where $z$ is the coordination number of the two species, $k_B$ is Boltzmann's constant, $T$ is the system temperature, $x_V$ and $x_L$ are the fraction of particles that are vacancies and lithium atoms, respectively (such that $x_V+x_L=1$), and the $\epsilon_{ij}$ terms are the interaction energies between species. According to this model, a phase transition occurs at a temperature
\begin{equation}
    T_c = \frac{\Omega}{2k_B},
\end{equation}
where $\Omega=\left(\nicefrac{z}{2}\right)\left(2\epsilon_{VL} - \epsilon_{VV} - \epsilon_{LL}\right)$. At temperatures higher than $T_c$, the vacancies are fully soluble in the mixture, and, at lower temperatures, two phases coexist: one with high concentration while the other with a low concentration of vacancies. In this formalism, a high vacancy phase corresponds to the onset of void formation.

\begin{figure}
    \centering
    \includegraphics[width=0.75\textwidth]{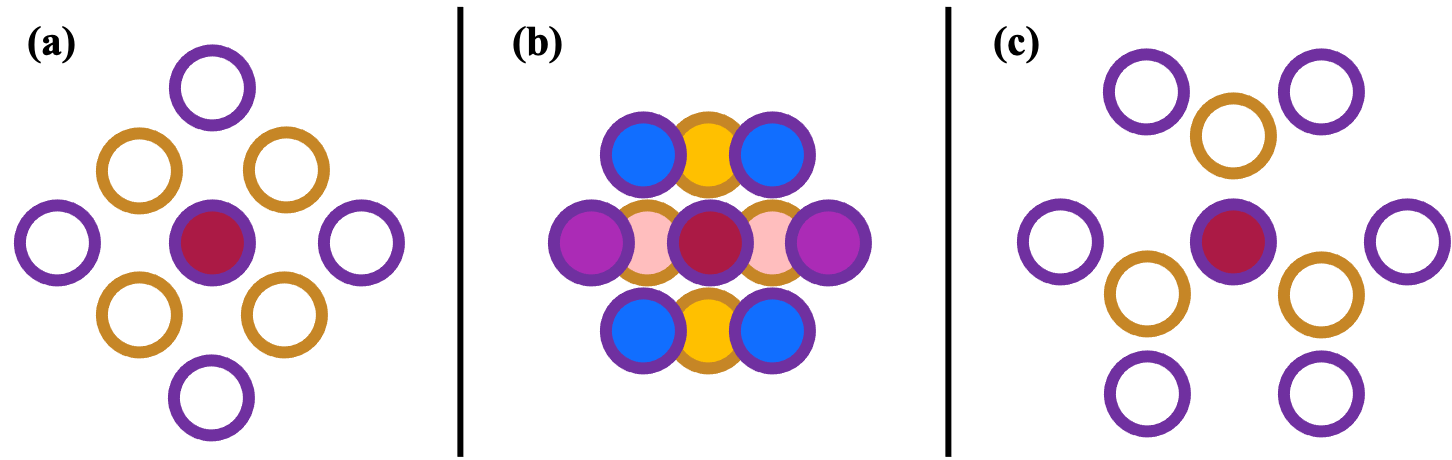}
    \caption{Top-view schematic of the different possible interactions considered for the \textbf{(a)} (100), \textbf{(b)} (110), and \textbf{(c)} (111) lithium surface facets. Purple circles denote lattice sites on the outer-most surface layer, and yellow orange circles denote sites on the second outer-most layer. In the (110) case (panel \textbf{(b)}), there are several different pair-wise interactions: red-purple, red-blue, red-yellow, and red-pink. In the other two cases, all purple-purple and orange-orange site interactions are equivalent.}
    \label{fig:surf_inter}
\end{figure}

Using density functional theory calculations, we analyze this model for three low Miller index lithium surface facets: (100), (110), and (111). A schematic of the different surface interactions considered is shown in Figure \ref{fig:surf_inter}. For each facet, we examined two possible interactions: one between vacancies which are both located on the same Miller plane on the surface (shown in purple), and another between vacancies that are located on different Miller planes, where one is at the surface, and another is right below it (shown in orange). For simplicity, we will use superscripts 1 and 2 to denote the surface-level Miller plane and the plane immediately under it, respectively. For example, $\Omega^{(11)}$ denotes the interaction parameter on the surface plane (intra-plane), while $\Omega^{(12)}$ corresponds to the interaction parameter between the two different Miller planes mentioned (inter-plane). Note that, for the (110) facet, further distinctions have to be made: there are two different intra-plane interactions, one with four neighbors, and one with two neighbors, as well as two different inter-plane interactions, both with only two neighbors, as represented in Figure \ref{fig:surf_inter}. Due to the limitations of the regular solution model, it is infeasible to take into account all these different possibilities separately. Therefore, in the case of the (110) surface, we incorporate these distinct interactions by averaging them together, in proportion to the number of neighbors involved.

We used density functional theory (DFT), coupled with the Perdew, Burke, and Ernzerhof (PBE) exchange correlation functional~\cite{GGAPBE} in the projector augmented wave (PAW) code GPAW~\cite{enkovaara2010electronic} implemented in the Atomic Simulation Environment (ASE)~\cite{larsen2017atomic}, to calculate the different interaction parameters. Computational details can be found in the Supplementary Information. The results of our analysis are summarized in Table \ref{tab:param_res}, which also includes the average distances, denoted by $d^{(11)}$ and $d^{(12)}$, between a lithium atom in the surface-level Miller plane and a neighboring lattice site, in either the same plane or in the one below. The Brillouin zone was sampled using the Monkhorst Pack scheme.

\begin{table}[htbp]
    \centering
    \begin{tabular}{c|ccc}
        Facet & (100) & (110) & (111) \\
        \hline \hline
        $z^{(11)}$ & 4 & 6 & 6 \\
        \hline
        $z^{(12)}$ & 4 & 4 & 3 \\
        \hline
        $\gamma$ [eV/\AA$^2$] & $\sim 0.030 $ & $\sim 0.031 $ & $\sim 0.034 $ \\
        \hline
        $\langle d^{(11)}\rangle$ [\AA] & 3.44 & 3.13 & 4.86 \\
        \hline
        $\langle d^{(12)}\rangle$ [\AA] & 3.03 & 3.17 & 3.05\\
        \hline
        $\Omega^{(11)} $ [eV] & 0.331 & 0.269 & 0.016 \\
        \hline
        $\Omega^{(12)}$ [eV] & 0.223 & 0.012 & -0.194 \\
        \hline
        $T^{(11)}_c$ [K] & $\sim 1980 $ & $\sim 1600 $ & $\sim 100 $ \\
        \hline
        $T^{(12)}_c$ [K] & $\sim 1340 $ & $\sim 70 $ & 0\\
        \hline
    \end{tabular}
    \caption{Approximate surface energies, coordination numbers, interaction parameters, and critical temperatures for different surface facets. Given the relatively small difference in surface energies, it is fair to assume all facets are equally relevant. The strongest interactions in the (100) and (110) facets have very high critical temperatures for vacancy solubility (above 1000 K), meaning that, under standard operating conditions, vacancies will tend to phase separate and begin forming voids. That is not the case for the (111) surface, where vacancies are fully solvable and, under equilibrium conditions, should be uniformly distributed on the surface.}
    \label{tab:param_res}
\end{table}

\begin{figure}
    \centering
    \includegraphics[width=0.75\textwidth]{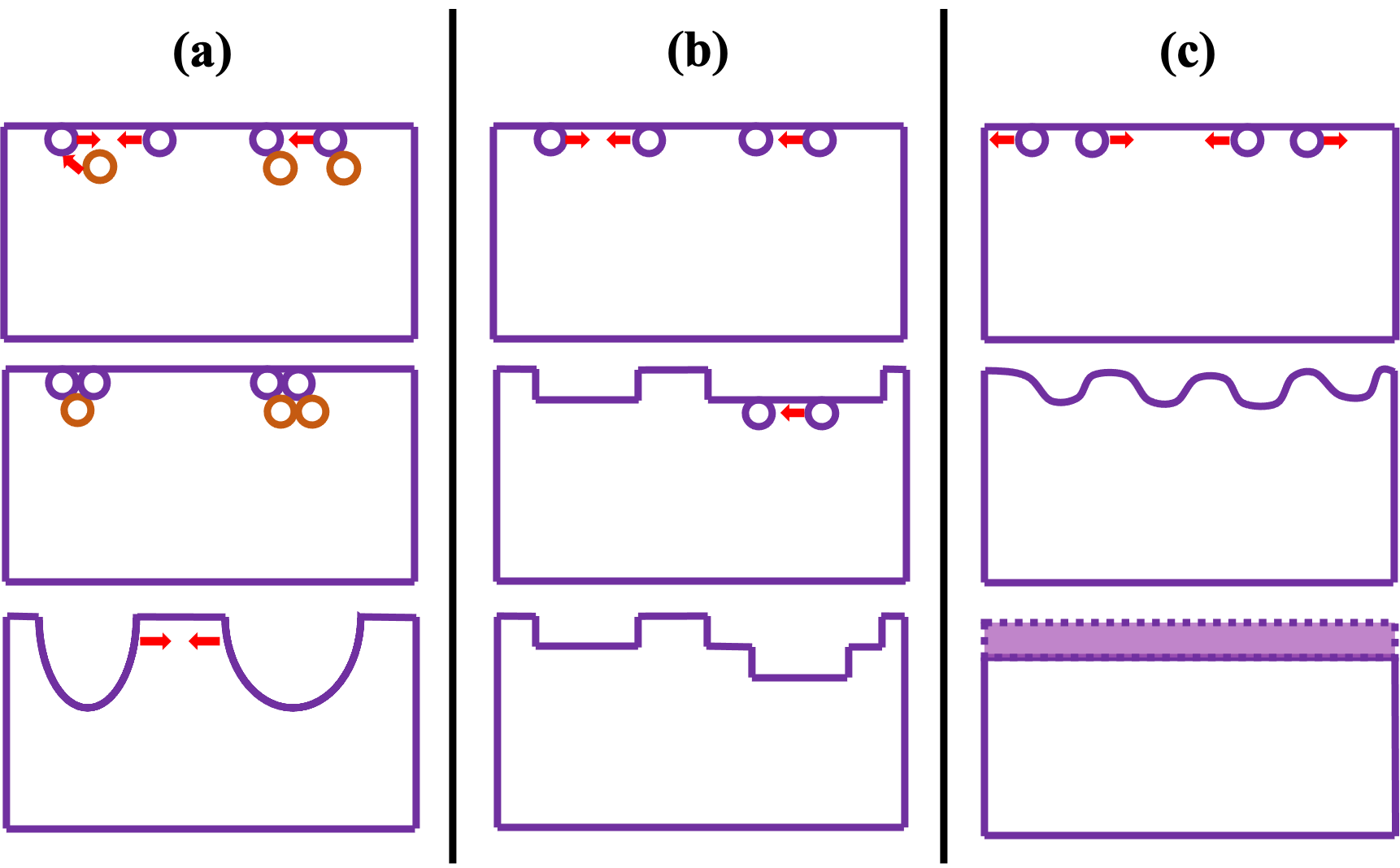}
    \caption{Side-view schematic of the expected behavior of  \textbf{(a)} (100), \textbf{(b)} (110), and \textbf{(c)} (111) lithium surface facets during stripping. The removal of atoms during discharge creates vacancies on the surface, denoted by circles. Purple circles represent vacancies on the surface-level Miller plane, and orange circles are vacancies on the second outer-most layer. Red arrows show the interactions between vacancies. In the (100) facet, vacancies will attract each other regardless of layer, and can thus form large (in an atomistic scale) valleys. The (110) surface exhibits a somewhat similar behavior, but, due to the inter-layer vacancy repulsion, vacancies would only congregate on the surface-level Miller plane. By doing so, they form steps that expose the underlying layer and creates a new surface for lithium extraction. In this case, valleys would be formed as a collection of terraces. The (111) facet shows the most promise: the strong inter-layer vacancy repulsion, coupled with the weak intra-layer attraction, will force vacancies to be uniformly distributed on the surface. In doing so, when enough lithium is stripped, a relatively flat surface will emerge. The light pink rectangle in panel (c) enclosed by the dashed line indicates the stripped layer.}
    \label{fig:discharge_schem}
\end{figure}

Our estimates indicate that all three facets have similar surface energy and hence, could all be formed during the electrodeposition process.  Controlling the facet has been demonstrated in the context of lithium electrodeposition~\cite{shi2017strong}. Vacancy-vacancy interactions are highly favorable in the (100) facet, regardless of the Miller planes of the vacancies, but their attraction is stronger when they are both on the top plane. This implies that the critical temperature for vacancy solubility in this surface is extremely high, above 1300 K, and that, under equilibrium conditions, vacancies will tend to phase separate and, thus, initiate the formation of large voids. Similar to the (100) facet, the strongest interactions in the (110) facet are also intra-plane; however, in this case, the inter-plane interactions are much weaker, and entropic contributions push vacancies in different planes apart even at low temperatures. Therefore, in this case, vacancies are expected to congregate on the top Miller plane, and, in doing so, form steps on the surface, which leads to increased surface roughness and facilitates void and pit formation. Interestingly, the (111) facet exhibits the opposite behavior: the intra-plane interactions, albeit attractive, are relatively weak, while the inter-plane ones are stronger, but highly repulsive. This suggests that, under equilibrium and at normal operating conditions, the (111) surface, by having perfect vacancy solubility, is the most likely to prevent the issue of void formation and pitting. The expected behavior of these surface is shown in Figure \ref{fig:discharge_schem}.

Note that, for all facets considered, vacancies on the same layer attract each other. However, these interactions are much stronger in the (100) and (110) cases, the two surfaces in which the distance between neighboring sites is closer to 3.5 \AA, the lattice parameter of a conventional lithium unit cell. For the (111) facet, the large distance $\langle d^{(11)}\rangle$ between neighboring sites diminishes the effects of a vacancy on its neighbors, allowing entropy to dominate the Gibbs free energy at even low temperatures. This large distance also contributes to the relatively high atomic-level roughness of this facet when compared to its (100) and (110) counterparts, something that is reflected in the relative surface energies of these facets. We believe that this partially contributes to the repulsive inter-layer interactions between vacancies: having neighboring vacancies in the two top layers would only increase said roughness. 

This work uses the regular solution model to examine vacancy congregation on different lithium metal surface facets, which constitutes the first step in the process of void formation and subsequent pitting that plagues lithium metal anodes. Our study identifies that the (111) surface facet could help mitigate this issue under equilibrium conditions.  Kinetic effects can promote vacancy congregation in the (111) facet: if vacancies are created faster than they can diffuse on the surface, voids, and eventually pits, will irrevocably form. This will be the subject of future investigations and will provide a deeper understanding of mechanism of void formation and pitting in lithium metal anodes.  This fundamental mechanistic understanding developed here provides a rational basis for enabling fast-discharging lithium metal batteries for electric aircraft.

\begin{acknowledgement}
This work was supported in part by the Advanced Research Projects Agency-Energy Integration and Optimization of Novel Ion Conducting Solids (IONICS) program under Grant No. DEAR0000774. Acknowledgment is also made to the Extreme Science and Engineering Discovery Environment (XSEDE) for providing computational resources through Award No. TG-CTS180061.
\end{acknowledgement}

\begin{suppinfo}
The following files are available free of charge.
\begin{itemize}
\item Li\_stripping\_surf\_SI.pdf: Computational details of simulations and of parameter estimation for regular solution model.

 \end{itemize}

 \end{suppinfo}


\providecommand{\latin}[1]{#1}
\makeatletter
\providecommand{\doi}
  {\begingroup\let\do\@makeother\dospecials
  \catcode`\{=1 \catcode`\}=2 \doi@aux}
\providecommand{\doi@aux}[1]{\endgroup\texttt{#1}}
\makeatother
\providecommand*\mcitethebibliography{\thebibliography}
\csname @ifundefined\endcsname{endmcitethebibliography}
  {\let\endmcitethebibliography\endthebibliography}{}

\includepdf[pages={1-3}]{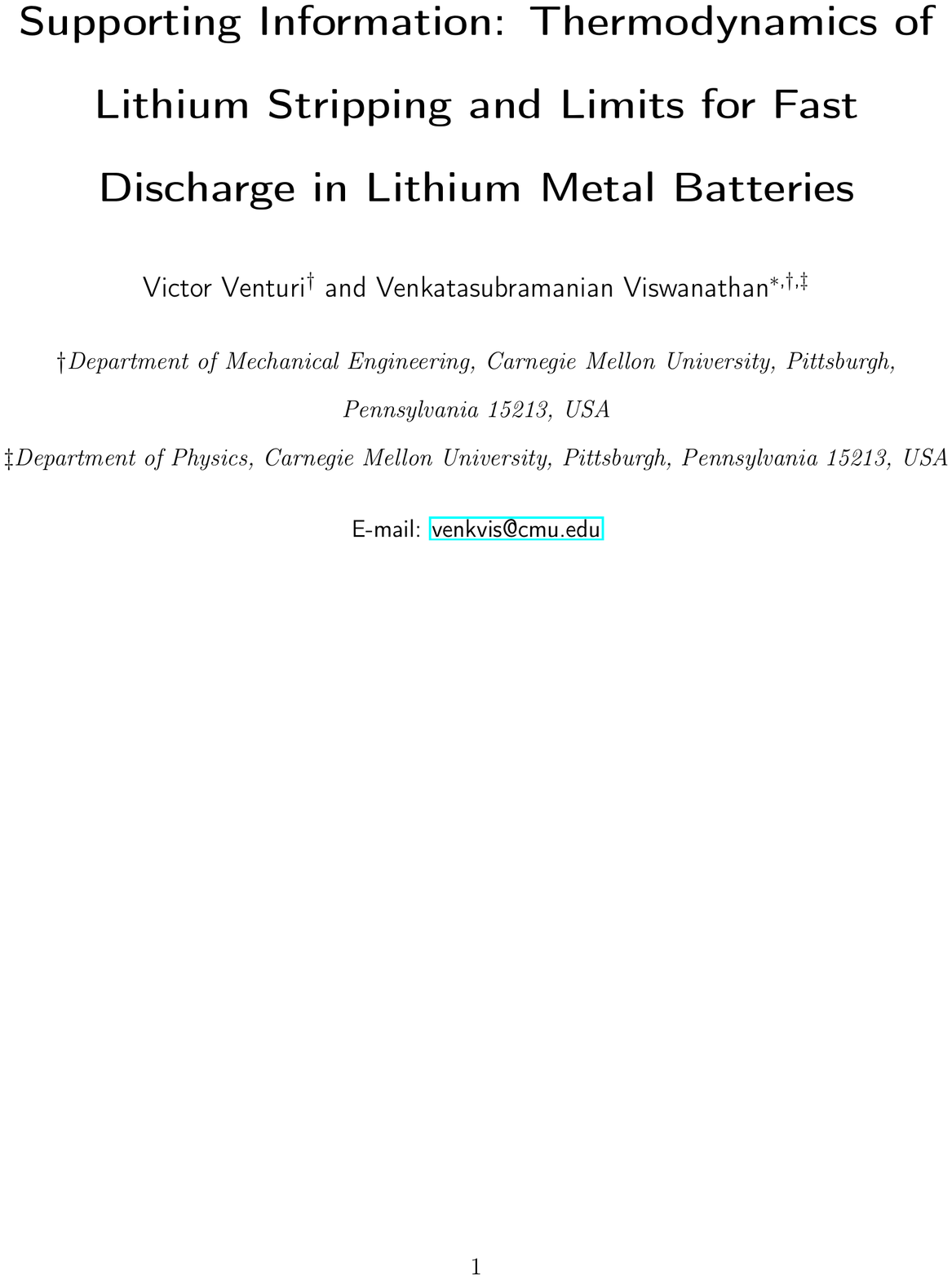}

\end{document}